*Hypothesis*

# Electron spin and the origin of Bio-homochirality II. Prebiotic inorganic-organic reaction model

Wei Wang

CCMST, Academy of Fundamental and Interdisciplinary Sciences, Harbin Institute of Technology, Harbin, 150080, China ; Email: wwang_ol@hite.edu.cn; Phone: 86-451-86418430; Fax: 86-451-86418440

**Abstract**

The emergence of biomolecular homochirality is a critically important question about life phenomenon and the origins of life. In a previous paper (arXiv:1309.1229), I tentatively put forward a new hypothesis that the emergence of a single chiral form of biomolecules in living organisms is specifically determined by the electron spin state during their enzyme-catalyzed synthesis processes. However, how a homochirality world of biomolecules could have formed in the absence of enzymatic networks before the origins of life remains unanswered. Here I discussed the electron spin properties in $Fe_3S_4$, ZnS, and transition metal doped dilute magnetic ZnS, and their possible roles in the prebiotic synthesis of chiral molecules. Since the existence of these minerals in hydrothermal vent systems is matter of fact, the suggested prebiotic inorganic-organic reaction model, if can be experimentally demonstrated, may help explain where and how life originated on early Earth.

## 1. Background

Symmetry is a fundamental aspect of nature. For example, certain molecules can occur in two enantiomeric forms. Common chemical synthesis always produce equal amount of these two forms. In living systems, however, this symmetry is broken. Naturally occurring proteins are composed of L-amino acids but no D-forms, whereas RNA and DNA contain D-form ribose and deoxyribose. Since the first discovery of molecular chirality by Pasteur in 1848 [1], the reasons behind those observations still remain a giant puzzle.

In a previous paper [2], I put forward a hypothesis that the emergence of a single chiral form of biomolecules in living organisms is specifically determined by the electron spin state during their enzyme-catalyzed synthesis. In the enzyme-promoted reaction, electrons released from a coenzyme, e.g., NAD(P)H, are heterogeneous in spin states; however, when passing through the chiral α-helix structure of the enzyme molecule to the other end of the helix, they experience the electrostatic potential and the inherent magnetic field of the molecule; their spin states are filtered and polarized, producing only "spin up" or "spin down" electrons; once the spin-polarized electrons participate a reductive reaction for the synthesis of a biomolecule, only L- or D-enantiomeric configuration is formed.

Then how could a homochirality world of biomolecules have formed in the absence of enzymatic networks before the origins of life? In this paper, I will talk about the spin properties of electrons in transition metal sulfides, such as ferromagnetic $Fe_3S_4$, paramagnetic ZnS and doped diluted magnetic ZnS, and their possible roles in the prebiotic synthesis of chiral molecules, based on the hydrothermal vent theory for the origins of life.

## 2. Hydrothermal vents and the origins of life

A hydrothermal vent is a fissure in a planet's surface from which geothermally heated water issues. At the bottom of an ocean, seawater infiltrates deep down into the earth's crust through cracks. After being heated by magma chambers, newly-born heat ocean crust, or the thermal energy released by serpentinization processes, the water with temperature up to 400 ℃ seeps through the crust, experiences complex water rock interactions (WSI), and then rises back into the ocean, carrying a lot of organic and inorganic components. Due to the high hydrostatic pressure (tens of millions of pascals) at the bottom of the ocean, water may exist in either its liquid form or as a supercritical fluid at such high temperatures. During the WSI, metals and hydrogen sulfide get dissolved out of the rock. The thermal fluid rises and gushes out of the rock as hot vent. When the fluid mixes with the cold seawater，metal sulfides like iron sulfide precipitate, forming chimney structures. The gushing thermal fluid and black precipitates are vividly called "black smoker".

Since the discovery of the first submarine hydrothermal vent in 1977 [3], it has been frequently argued as a congenial site for the origin of terrestrial life [4-8]. This scenario makes sense and rapidly gained ground because many primitive organisms were found near the vents [9]. In fact, this discovery is only one of several reasons why in the last thirty five years interest has been focused on hydrothermal vents in general as potential sites for life's origin. Some other aspects are listed as follows: 1) the high temperature and high pressure supercritical fluids containing a large amount of organic and inorganic components help the prebiotic synthesis of biomolecuels [10-12].

2）The minerals could have acted as a catalyst, a template and a power source for the prebiotic synthesis of biomolecules, the crucial evolutionary nascence of ancient metabolic pathways, and the emergence of the first cells (vesicles) [5, 7, 13]. 3) The hydrothermal systems are about the only niche where primitive life would have been protected against extensive meteorite impacts on early Earth and partial vaporization of the ocean [12]. 4) In 2000, Rasmussen [14] reported a discovery of pyritic filaments in a 3,235-million-year-old deep-sea volcanogenic massive sulfide deposit from the Pilbara Craton of Australia. The fossil remains of thread-like microorganisms which were probably thermophilic chemotropic prokaryotes. 5) In all extant life-forms, there exists a family of iron-sulfur proteins [15, 16]. The Fe-S clusters at the active centers of those proteins are similar to some sulfide minerals. For instance, the $Fe_4S_4$ 'thiocubane' unit in carbon monoxide dehydrogenase and ferredoxins looks very like the $Fe_4S_4$ 'cubane' unit of greigite, $Fe_3S_4$ [7, 17, 18]. Therefore, a notion has been entertained that iron-sulfur proteins may evolve from those mineral structures sequestered by primordial abiotic short peptides, and the whole to act as the first electron transfer agents, hydrogenases and synthetases [19].

Among the various sulfide minerals composed black smoker chimneys, the original major precipitate is amorphous FeS [5], which may further evolve into mackinawite, greigite, pyrrhotite, and ultimately end up mainly as pyrite. In the presence of certain organic molecules, however, the amorphous FeS can be selectively transformed into greigite, a ferromagnetic mineral [20]. Both these two minerals may have played roles of key importance in prebiotic chemistry [5, 7].

ZnS is another concernful component of submarine chimneys. Among all metal sulfide species therein, ZnS precipitates more slowly [21]. In addition, Zn can be stripped and remobilized from buried chimney fragments, and concentrated at the seawater interface due to some hydrothermal geological activities [22]. Therefore, ZnS is a prevalent constituent at the periphery of hydrothermal deposits, well known as sphalerite. Zinc sulfide (ZnS) is an n-type semiconductor characterized by a wide bandgap of about 3.6 eV. It can absorb UV light with a wavelength shorter than 344 nm to produce photoelectron/hole pairs and catalyze redox reactions [23]. The reaction of the photoelectron with an adsorbate leads to the reduction of the adsorbate, while the hole induces oxidation reactions. This property of ZnS could not only make possible the carbon/nitrogen fixation [24, 25] and molecular chain extension processes [26-28] to form organic molecules, but also help maintain the redox homeostasis between certain biomolecules [18, 23, 29, 30]. If the hydrothermal vents on early Earth were located at a shallow depth no more than 100 m which consisted of the "photic zone" of primitive ocean, sunlight could penetrate the waters of early Earth to trigger prebiotic synthesis on ZnS surfaces [29, 30]. In hydrothermal fluids, Fe, Ni, and Cu are more common than Zn. They always occur in the form of pyrite ($FeS_2$), pentlandite

((Fe,Ni)$_9$S$_8$), chalcopyrite (CuFeS$_2$), etc [31, 32]. In early hydrothermal vents, whether during the formation of pristine ZnS or its remobilization and reprecipitation at the seawater interface, these transition metal ions may co-deposit with ZnS, forming new energy levels in the band structure of ZnS. This doped ZnS may not only response to longer wavelength light (> 344 nm, even visible light) [33, 34], but also from dilute magnetic semiconductor [35-38].

## 3. Electron spin properties in ferromagnetic Fe$_3$S$_4$, paramagnetic ZnS and doped dilute magnetic ZnS

3.1 Greigite (Fe$_3$S$_4$)

Greigite (Fe$_3$S$_4$) is an iron thiospinel, which has the same crystal structure as magnetite (Fe$_3$O$_4$) and crystallizes in the inverse spinel structure. Greigite is ferrimagnetic, with the spin magnetic moments of the Fe cations in the tetrahedral sites oriented in the opposite direction to those in the octahedral sites, and a net magnetization. It is a mixed-valence compound, featuring both Fe$^{2+}$ and Fe$^{3+}$ centers in a 1:2 ratio. Both metal sites have high spin quantum numbers. The electronic structure of greigite is that of a half metal [39].

Half metal is an extreme case of completely spin polarized material because its electronic state is metallic for one spin direction and insulating for the other. In other words, a half-metal acts as a conductor to electrons of one spin orientation, but as an insulator or semiconductor to those of the opposite orientation. This results in conducting behavior for only electrons in the first spin orientation. If the magnetization of the materials is reversed, the spin direction also reverses. Thus, depending on the direction of magnetization of a material relative to the spin polarization of the current, a material can function as either a conductor or an insulator for electrons of a specific spin polarization.

The spin polarization is defined as the ratio of the density of states of up-spin and down-spin electrons at a Fermi level, $P=(D_\uparrow-D_\downarrow)/(D_\uparrow+D_\downarrow)$, as shown in Fig. 1. For normal metals or paramagnetic materials, P=0 since the densities of states of up-spin and down-spin electrons are equal. For ferromagnetic materials such as Fe and Co, P is larger than 0 but smaller than 1 because the density of state of up-spin and down-spins are different. However, if a material has a band gap in the minority band at a Fermi level and exhibits metallic behavior in the majority band, the density of state of the minority band is zero at the Fermi level. A half-meal is of this case, and only up-spin electrons are present at the Fermi level, that is P=1.

So the mineral greigite can act as a spin filter under the action of an external magnetic field to produce spin polarized electrons in a single orientation and to induce the synthesis of chiral organic molecules.

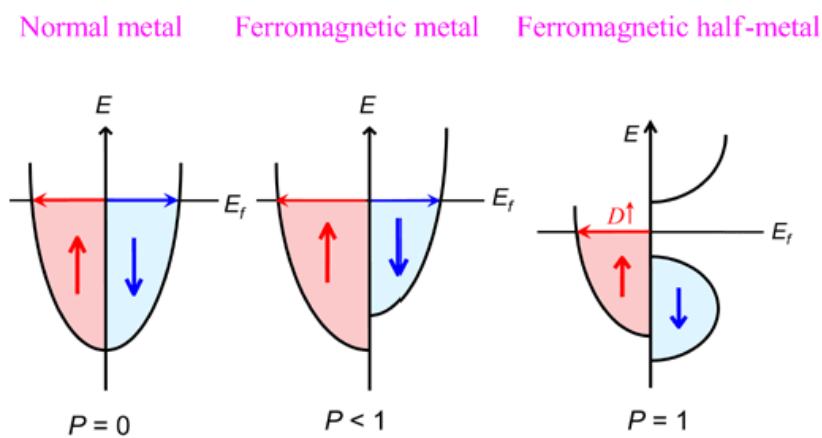

**Fig. 1** Electronic structure of paramagnetic, ferromagnetic and ferromagnetic half-metal materials.

Later, in the evolutionary context, $Fe_3S_4$ would have reacted with prebiotic peptides and finally evolved into extant Fe-S proteins, whose Fe-S cluster active centers are strikingly similar to the iron sulfide minerals in tiny structure [7] and then act as a spin filter to induce the asymmetric synthesis of chiral biomolecules. However, the cofactors of extant amino acid dehydrogenases, which catalyze the synthesis of amino acids, are nicotinamide adenine dinucleotide NAD(P)H but not Fe-S cluster [2]. This seeming contradiction does not threaten the suggested model, since in evolution the latter might have largely been replaced by NAD(P)H, as shown in the Entner-Doudoroff pathway of *P. furiosus*, which belongs to the most primitive organisms [18, 40]. Therefore, I would expect that the enzymes for stereoselective synthesis in extant organisms may be evolutionary vestiges of a primordial inorganic ancestor.

3.2 ZnS and doped ZnS dilute magnetic semiconductor

As mentioned above, zinc sulfide (ZnS) is an n-type semiconductor. It can absorb UV light with a wavelength shorter than 344 nm to produce conduction band electrons and valence band holes. On the other hand, photons of right or left polarized light have a projection of the angular momentum on the direction of their propagation. When a circularly polarized photon is absorbed by ZnS, this angular momentum is distributed between the photo-excited electrons and holes according to the selection rules determined by the band structure of the semiconductor (Fig. 2). In nature, there exist several sources of circularly polarized light [41-46]. They may penetrate the waters of early Earth, reach the surfaces of sphalerite, produce spin polarized electrons or holes in ZnS particles, and then trigger prebiotic asymmetric synthesis (Fig. 2).

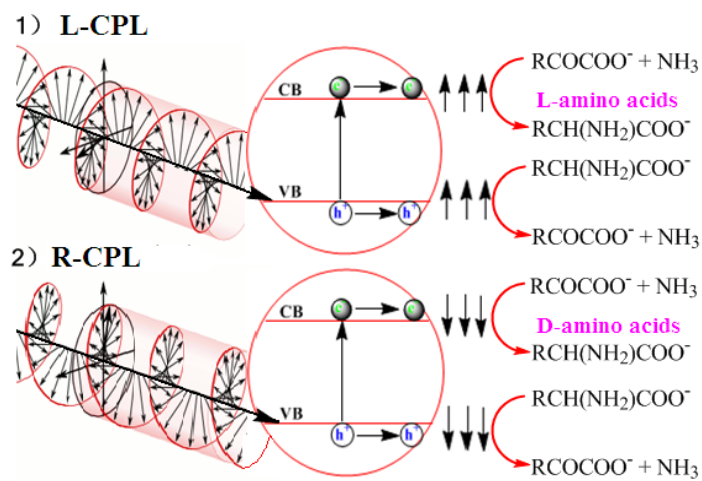

**Fig. 2** Proposed reaction scheme diagram for the asymmetric synthesis or degradation of amino acids on L- or D-circularly polarized light (CPL) illuminated ZnS surfaces. CB, conduction band; VB, valence band.

ZnS is a paramagnetic II-VI semiconductor material. The density of states of up-spin electrons equals to that of down-spin electrons. However, when it is doped by magnetic transition metal, the exchange interaction of electrons with magnetic ions leads to the giant Zeeman spin splitting of the conduction band. As a result, the electron densities in the spin-up and spin-down subbands become different [47-49], just like the case of ferromagnetic metal (Fig. 1). A net spin-polarized electric current can be obtained, even by linearly polarized light excitation [50, 51] or microwave radiation [52].

## 4. Validation: Additional tests of the suggested model

According to the above arguments, $Fe_3S_4$, ZnS, and transition metal doped dilute magnetic ZnS can produce net spin polarized electric currents under the action of a magnetic or light field. This symmetry-broken physical factor might have participated the prebiotic asymmetric synthesis of biomolecules.

To test the suggested hypothesis, one can prepare $Fe_3S_4$, ZnS, or dilute magnetic ZnS to catalyze abiotic syntheses of chiral molecules (for example, the synthesis of α-amino acids [23, 29, 30]), by using electrochemical or photochemical methods. In some cases, an external magnetic field is needed for the regulation of electron spin. The sulfide minerals should be prepared in nanosize since at room temperature the spin relaxation time and spin diffusion length of different materials is only about hundreds of nanoseconds and several to hundreds of nanometers, respectively.

## 5. Conclusions

In this paper, I discussed the electron spin properties of $Fe_3S_4$, ZnS, and transition metal doped dilute magnetic ZnS, and their possible roles in the prebiotic synthesis of chiral molecules. Since the existence of these minerals in hydrothermal vent systems is matter of fact, the suggested prebiotic inorganic-organic reaction model, if can be experimentally demonstrated, may give a support to the hydrothermal vent theory for the origins of life on early Earth. Life is just like a lone boat floating on a boundless sea. We don't know where we came from. The suggested model may tell us the answer, just like a light tower in the darkness.